\documentclass[11pt,twocolumn]{article} % twocolumn 옵션 추가
\usepackage[utf8]{inputenc}
\usepackage{geometry}
\usepackage{graphicx}
\usepackage{caption}
\usepackage{multicol}
\usepackage{titlesec}

\titleformat{\section}{\normalfont\large\bfseries}{\thesection}{1em}{}
\titleformat{\subsection}{\normalfont\normalsize\bfseries}{\thesubsection}{1em}{}

\title{\textbf{Advanced 3D Imaging Approach to TSV/TGV \\ Metrology and Inspection Using Only Optical Microscopy}}
\author{Gugyeong Sung}
\date{}

\begin{document}

% Abstract만 단일 칼럼
\twocolumn[
\maketitle

\begin{abstract}
This paper introduces an innovative approach to silicon and glass via inspection, which combines hybrid field microscopy with photometric stereo. Conventional optical microscopy techniques are generally limited to superficial inspections and struggle to effectively visualize the internal structures of silicon and glass vias. By utilizing various lighting conditions for 3D reconstruction, the proposed method surpasses these limitations. By integrating photometric stereo to the traditional optical microscopy, the proposed method not only enhances the capability to detect micro-scale defects but also provides a detailed visualization of depth and edge abnormality, which are typically not visible with conventional optical microscopy inspection. The experimental results demonstrated that the proposed method effectively captures intricate surface details and internal structures. Quantitative comparisons between the reconstructed models and actual measurements present the capability of the proposed method to significantly improve silicon and glass via inspection process. As a result, the proposed method achieves enhanced cost-effectiveness while maintaining high accuracy and repeatability, suggesting substantial advancements in silicon and glass via inspection techniques.\\ \textbf{Keywords: TSV; TGV; 3D reconstruction; optical microscopy; photometric stereo; wafer inspection} 
\end{abstract}
\vspace{1em}
]

\section{Introduction}
\indent{}\indent{} With the development of semiconductor technology for artificial intelligence, high performance computing and wireless connectivity, the requirement on three dimensional integrated circuits (3D ICs) and high-density interconnect circuit boards (HDIs) is increasing [1-3]. These essential technologies are implemented by silicon vias (TSVs) and through glass vias (TGVs) to realize chip stacking and higher integration for enhanced performance [2,4,5]. Silicon wafers are the most popular material for semiconductor fabrication owing to their electrical suitability and economy, and glass wafers offer thermal stability, chemical resistance and lower parasitic capacitance and hence are also gradually finding their place in the advancement of next generation semiconductor technologies [6-9].

Traditional via defects are usually inspected by scanning electron microscopy (SEM) or X-ray imaging as conventional inspection method [10-13]. However, there are some restrictions; low inspection speed, the requirement for high vacuum environment, and the high-cost maintenance and safety features such as radiation shielding. For these challenges, they are typically employed as post-processing evaluation tools and cannot be used for in-line or online examinations in high volume production. This dependency on traditional techniques requires faster, cost-effective, and safer inspection methods to guarantee the structural integrity and functionality of vias.

To overcome these limitations, this study proposes a photometric stereo-based inspection method for silicon and glass that utilizes their specific optical characteristics. Photometric stereo is a technique that uses captured images under different lighting conditions, and the proposed method, the photometric stereo based optical microscopy, enables to reconstruct the 3D shape of TSV/TGV structures to analyze surface texts and internal variations. This method allows non-invasive inspection in the visible and near infrared ranges using the silicon and glass transmittance to image the interior of the vias and their surroundings without requiring expensive preparation steps. Therefore, the proposed method offers an accurate, economical, and fast inspection method of detecting micro defects for the growing need for online or at line inspection in semiconductor production facilities.

\begin{figure}[t]
\centering
\includegraphics[width=\linewidth]{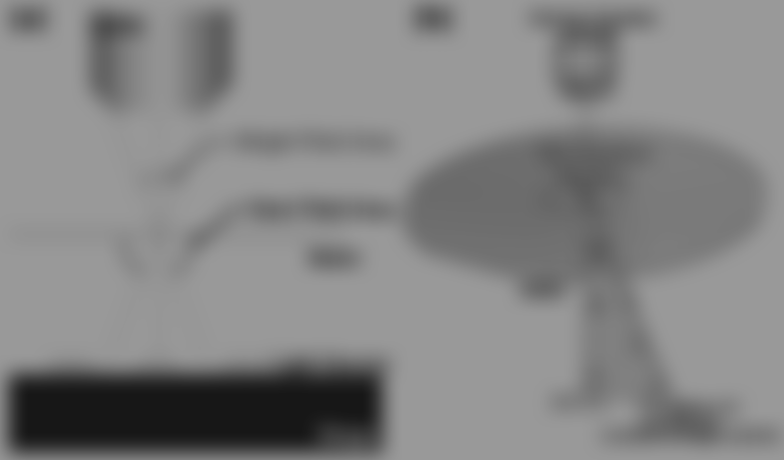}
\caption{A SCHEMATIC OF (A) THE HYBRID FIELD MICROSCOPY WITH (B) PHOTOMETRIC STEREO-BASED PROPOSED METHOD}
\label{Fig1}
\end{figure}

\section{MATERIALS AND METHODS}\vspace{-1.2em}
\subsection{Traditional Inspection Method}
\indent{}\indent{} Via inspection in substrates is generally performed using optical microscopy, SEM, and X-ray microscopy to analyze structural integrity, detect defects, and evaluate dimensional accuracy [10-12]. Optical microscopy is widely used due to its cost-effectiveness, simplicity, and non-destructive method. Moreover, dark field (D-field) microscopy particularly is efficient in the detection of micro defects such as cracks and particles. However, optical microscopy is limited to surface inspection, and detecting subsurface defects typically requires special preparation. To overcome some of these restrictions, through-focus scanning optical microscopy (TSOM) has been developed to produce 2D images at different focal positions and construct the 3D model, which enables the non-destructive technique [12,14]. While TSOM effectively leverages standard optical microscopes as 3D metrology tools, it requires accurate focusing and a large number of images for high resolution, making the process time-intensive.

SEM and X-ray microscopy are generally utilized to improve conventional optical inspections. SEM provides high spatial resolution and depth of field for detailed surface analysis but has drawbacks of sputter coating, high vacuum, and high cost, and can lead to substrate damage by sectioning particularly for fragile materials like silicon and glass. X-ray microscopy enables non-destructive internal inspection of via features and thus is useful for determining the shape and alignment of features without the need for sectioning. However, it has a low spatial resolution that hinders microdefect detection, and the protective shielding required increases the cost of operation and system complexity [10–13]. SEM and X-ray microscopy resolve some limitations of optical microscopy. However, their high cost, long inspection time, and safety concerns create a need for efficient, on-line in real-time, and cost-effective inspection methods in semiconductor manufacturing.

\subsection{Proposed Inspection Method}
\indent{}\indent{} Conventional optical microscopy with bright field (B-field) and dark field (D-field) techniques are widely utilized for defect inspection due to its cost-effectiveness and simplicity. B-field microscopy provides high-resolution surface detail under direct illumination, and D-field microscopy enhances contrast, effectively highlighting particles, edges, and boundaries. However, both methods are limited that shallow depth of field makes them less effective for inspecting complex 3D structures of TSV/TGVs.

To address this limitation, this study proposes a hybrid inspection method that integrates photometric stereo with conventional optical microscopy. The photometric stereo technique captures multiple images under different lighting conditions to determine the surface normal directions and reconstructs 3D features. This technique reveals fine texture and structural details essential for identifying intricate features. As shown in Figure 1(a), the optimization of B-field and D-field microscopy areas can be achieved by utilizing the numerical aperture (NA) of the lens to enhance resolution and contrast by controlling the acceptance angle.

Therefore, the proposed method combines photometric stereo based techniques and the strengths of both B-field and D-field microscopy to obtain not only the high spatial resolution of surface features from B-field microscopy but also the improved edge contrast from D-field microscopy. This integrated strategy enables the detection of micropores and edges on vias that cannot be easily seen in conventional optical microscopy and reconstructs these features in 3D from 2D images.

\section{Methodology}\vspace{-1.2em}
\subsection{Photometric stereo}

\indent{}\indent{} Photometric stereo is a technique used to estimate 3D surface characteristics by capturing multiple images of an object under different lighting conditions with a fixed camera [15]. This approach analyzes changes in image intensity based on light source direction, enabling the calculation of surface normal by solving a system of linear equations. For surfaces with micro-scale roughness, such as fabricated vias, the Lambertian reflection model is often applied. By capturing at least three images with varying light directions, the surface normal and albedo at each pixel can be determined, with the normal providing critical data on surface orientation and partial derivatives of the depth map.

The depth map, representing the 3D surface, is achieved by integrating these partial derivatives and solving the Poisson equation. This process involves computing the Laplacian of the depth map and using the discrete cosine transform (DCT) to handle boundary conditions efficiently. The inverse DCT reconstructs the depth map in the frequency domain, ensuring accurate representation by removing any linear trends, so the depth map starts at zero, reflecting true surface topography. This method enables detailed 3D reconstruction, allowing for effective modeling of complex micro-textured surfaces.

\subsection{Hybrid Field Microscopy with Photometric Stereo}
\begin{figure}[t]
\centering
\includegraphics[width=\linewidth]{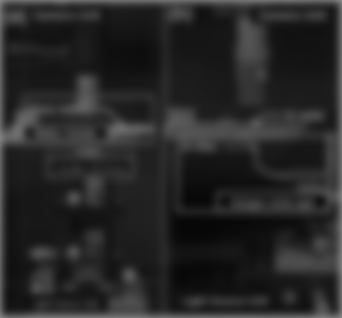}
\caption{OVERALL COMPOSITION OF EXPERIMENTAL EQUIPMENT FOR (A) TGV AND (B) TSV MEASUREMENTS}
\label{Fig2}
\end{figure}

\indent{}\indent{} For the hybrid field optical microscopy, light sources covered both B-field area and D-field areas, and the dark field area was induced by giving an angle of more than two times of objective aperture angle, $\underline{\theta}$, according to the NA value of the objective lens. Referring to the refractive index of silicon and glass substrates, it was mounted at the angle, which is lower than the critical angle, $\overline{\theta}$, of the light to avoid total reflection, as shown in Figure 1 (b). Therefore, the range of the height of light position can be defined as: 

\section{Experiment}\vspace{-1.2em}
\subsection{Experimental setup}

\indent{}\indent{} The entire system consists of a camera unit, a light source unit, and a wafer holder to demonstrate the proposed method, as shown in Figure 2. The camera unit comprises a camera (acA5472-17uc; BASLER AG), 20x objective lens, and c-mount microscope. The light source unit includes seven RGB LEDs on the micro-stage mounted beneath the glass via set, with the 20x objective lens securely aligned toward the center of the LED array, as shown in Figure 2(a). Additionally, a halogen white light source was used to apply the proposed method to TSVs, as the visible light spectrum cannot pass through silicon material.

To accommodate this, the camera (Alvium 1800 U-130 VSWIR; Allied Vision) and 50x NIR objective lens are configured to detect near-infrared (NIR) light spectrum to suit the specific requirements of silicon-based materials, as shown in Figure 2 (b).

\subsection{Data Acquisition and Analysis}

To ensure accuracy and repeatability in the proposed 3D reconstruction method, precise alignment was achieved between the camera, light source centers, and the sample, with careful adjustments to maintain the entire via set within the camera's field of view. For each blind TSV and TGV set, multiple images were captured under meticulously defined lighting conditions, and the light source positions were normalized into unit vectors for directional consistency. This directional consistency allowed accurate surface normal estimation by calculating each pixel’s surface normal and albedo through a least-squares approach, minimizing image noise and enhancing surface detail clarity. The calculated albedo further contributed to characterizing the surface's reflectance properties, supporting a comprehensive 3D reconstruction in MATLAB.

The surface normal data were converted into a 3D depth map using the Poisson equation, which translates directional information into physical height variations, revealing intricate depth and texture details in each via. To correct for depth discrepancies, especially around edges impacted by heat-affected zones and recast layers, a Gaussian filter was applied to smooth depth data based on proximity to the average depth. This leveling process assigned greater weight to values close to the average, reducing noise and discrepancies near edges and yielding a more accurate 3D structure. The depth maps were then quantitatively evaluated against reference values across multiple trials to ensure measurement consistency, confirming the reconstructed depth maps’ accuracy in representing the via structures.

\begin{figure}[t]
\centering
\includegraphics[width=\linewidth]{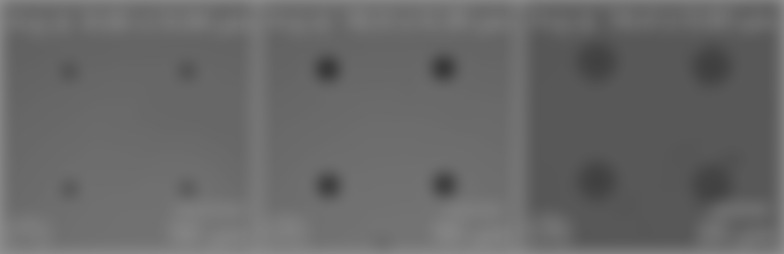}
\caption{2X2 ARRAY OF BLIND TSV/TGV SETS}
\label{Fig3}
\end{figure}

\section{RESULTS AND DISCUSSION}

\indent{}\indent{} In this experiment, 2x2 arrays of three sets of blind TSVs and TGVs were evaluated as shown in Figure 3. Figure 3 (1) and (2) display blind TSVs, fabricated using an etching process, and Figure 3 (3) shows a blind TGV fabricated by femtosecond laser. Depth maps for each via were reconstructed in 3D by collecting image data under five lighting conditions for TSVs and seven for TGVs. To enhance accuracy, depth discrepancies around via edges were corrected using a leveling process that assigned higher weights to depth values closer to the average depth, mitigating noise and achieving a smoother surface representation. The depth maps were quantitatively assessed by comparing measured depth values to known references, ensuring consistency and accuracy across multiple trials. The method also analyzed roundness errors in the vias using a least-squares circle (LSC) approach.

\begin{figure}[t]
\centering
\includegraphics[width=\linewidth]{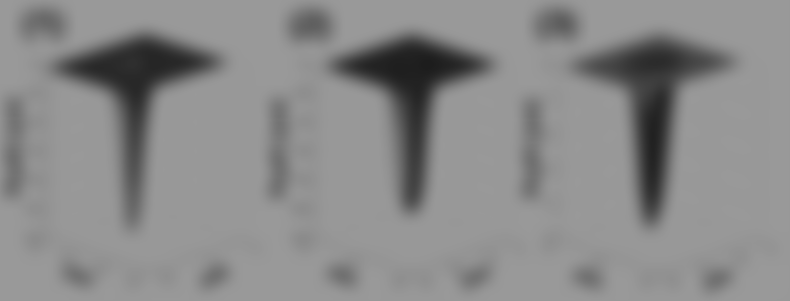}
\caption{3D DEPTH MAP RECONSTRUCTION USING THE PROPOSED METHOD}
\label{Fig4}
\end{figure}

Figure 4 demonstrates the visualized 3D depth maps of blind TSV/TGV, which is reconstructed to evaluate their structural characteristics. Samples (1) and (2) represent blind TSVs, which were fabricated using the etching process, while Sample (3) is a blind TGV, which was created through femtosecond laser processing. These depth maps provide intuitive insights into the structural differences in sub-layers.

The steep slopes seen in the reconstructed 3D depth maps are primarily a result of the images being squeezed horizontally during visualization, as shown in Figure 4. However, they’re also influenced by the limitations of the camera’s resolution. 

For example, the limitations of the resolution and the depth of focus of the camera system may not fully capture the finer details of the structural gradients. Therefore, these limitations may exaggerate the sidewalls of the via in the visualization.

Figure 5 demonstrates the comparison between the reference and measured samples using the proposed method, showing moderate differences in diameters and minor deviations in depth values within an acceptable range for inspection applications. While selective filtering was carefully applied to minimize discrepancies, the differences in diameter and depth can be attributed to unexpected light scattering and diffraction effects at the sharp via edges, which were not fully compensated for during reconstruction.

Figure 6 illustrates the roundness error trends across different depths in both normal and defective vias. Samples fabricated via the etching process exhibited consistently stable roundness with little variation throughout the depth. In contrast, the sample created through femtosecond laser processing showed greater variability in roundness, particularly near the top and bottom boundaries of the via. These deviations are likely due to process-induced surface irregularities and residuals, highlighting how fabrication techniques influence the geometric integrity of via structures.

Overall, the proposed method effectively reconstructs 3D structures of blind TSV/TGV with less volume of image data, providing both quantitative depth information and intuitive 3D representations. However, as the depth increases, the number of data points decreases due to the resolution limitations of the camera, which leads to a reduction in accuracy at greater depth. The lower resolution may produce slight discrepancies in estimating exact depth values and result in distortions near the bottom of the vias. Despite potential discrepancies from surface unevenness or camera adjustments, the method remains cost-effective and adaptable, making it highly valuable for detailed inspection of intricate via structures and reliable evaluation in complex 3D inspections.

\begin{figure}[t]
\centering
\includegraphics[width=\linewidth]{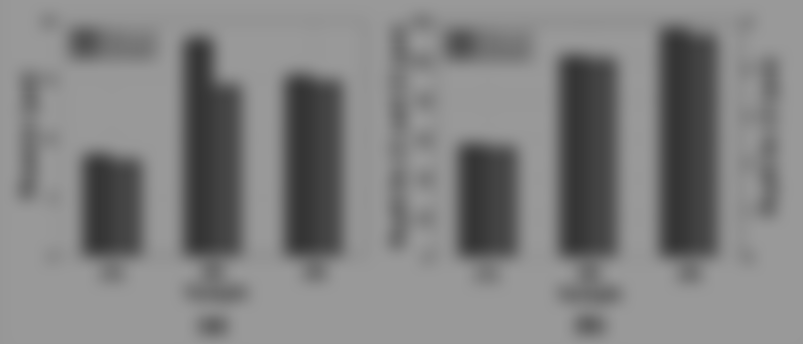}
\caption{COMPARISON OF (A) DIAMETERS AND (B) DEPTH BETWEEN THE REFERENCE AND ESTIMATED MODELS FOR BLIND TSV/TGV SETS}
\label{Fig5}
\end{figure}

\section{CONCLUSION}
\indent{}\indent{} This study introduces a novel approach combining hybrid field microscopy with photometric stereo for detailed 3D reconstruction of blind TSV/TGV using optical microscopy. The method effectively captures geometric and surface information, though challenges with light path accuracy in heat-affected zones remain. Offering quantitative 3D insights into blind TSV/TGV, this cost-effective technique holds strong potential for substrate and semiconductor applications. Future improvements will aim to enhance defect classification and identification in via manufacturing.

\begin{figure}[t]
\centering
\includegraphics[width=0.72\linewidth]{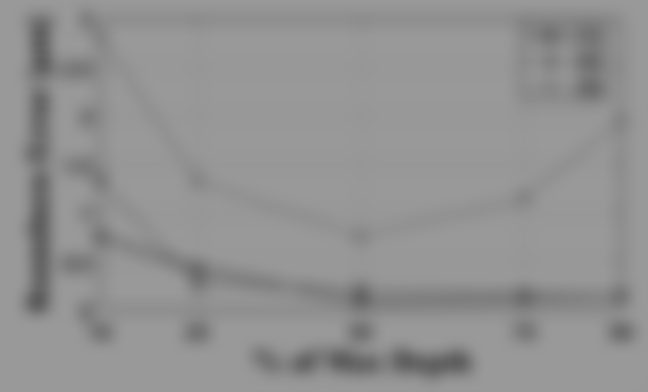}
\caption{ROUNDNESS EVALUATION ACROSS THE DEPTH USING LSC METHOD}
\label{Fig6}
\end{figure}

\section{FUTURE WORK}
\indent{}\indent{} I will focus on enhancing the accuracy and robustness of the proposed method with larger sample volume. Imaging resolution will be improved to mitigate the loss of data points at greater depths, reducing discrepancies in diameter and depth estimation. An advanced algorithm for tracking light direction will be applied to minimize systematic errors caused by various light conditions. Moreover, a larger sample volume for TSV/TGV sets will be incorporated to ensure a more comprehensive validation of sub-layer defect analysis. These improvements will contribute to a more precise and reliable approach for via inspection using the proposed method, which provides complex 3D structures using only optical microscopy with less volume of image data.

\section{ACKNOWLEDGEMENTS}
The research was supported by J.Mike Walker ’66 Department of Mechanical Engineering.

\section{REFERENCES}
[1] Zhang, K., 2024, "1.1 Semiconductor Industry: Present \& Future," 2024 IEEE International Solid-State Circuits Conference (ISSCC), Vol. 67, IEEE.
\newline
[2] Agonafer, D., 2015, “A Review of Cooling Road Maps for 3D Chip Packages,” Cooling of Microelectronic and Nanoelectronic Equipment: Advances and Emerging Research, pp. 1-17.
\newline
[3] Lau, J. H., 2011, "Evolution, Challenge, and Outlook of TSV, 3D IC Integration and 3D Silicon Integration," 2011 International Symposium on Advanced Packaging Materials (APM), pp. 462-488, IEEE.
\newline
[4] Lamy, Y., Dussopt, L., El Bouayadi, O., Ferrandon, C., Siligaris, A., Dehos, C., and Vincent, P., 2013, "A Compact 3D Silicon Interposer Package with Integrated Antenna for 60 GHz Wireless Applications," 2013 IEEE International 3D Systems Integration Conference (3DIC), pp. 1-6, IEEE.
\newline
[5] Rajmane, P., 2018, Multi-Physics Design Optimization of 2D and Advanced Heterogeneous 3D Integrated Circuits, Ph.D. Dissertation, The University of Texas at Arlington.
\newline
[6] Seok, B. C., and Jung, J. P., 2024, "Recent Progress of TGV Technology for High Performance Semiconductor Packaging," Journal of Welding and Joining, 42(2), p. 156.
\newline
[7] Ahmed, O., Jalilvand, G., Okoro, C., Pollard, S., and Jiang, T., 2020, "Micro-Compression of Freestanding Electroplated Copper Through-Glass Vias," IEEE Transactions on Device and Materials Reliability, 20(1), pp. 199-203.
\newline
[8] Yu, C., Wu, S., Zhong, Y., Xu, R., Yu, T., Zhao, J., and Yu, D., 2023, "Application of Through Glass Via (TGV) Technology for Sensors Manufacturing and Packaging," Sensors, 24(1), p. 171.
\newline
[9] Kim, J., Shenoy, R., Lai, K.-Y., and Kim, J., 2014, "High-Q 3D RF Solenoid Inductors in Glass," 2014 IEEE Radio Frequency Integrated Circuits Symposium, IEEE.
\newline
[10] Kudo, H., Takano, T., Akazawa, M., Yamada, S., Sakamoto, K., Kitayama, D., Iida, H., Tanaka, M., and Tai, T., 2021, "High-Speed, High-Density, and Highly-Manufacturable Cu-Filled Through-Glass-Via Channel (Cu Bridge) for Multi-Chiplet Systems," 2021 IEEE 71st Electronic Components and Technology Conference (ECTC), IEEE.
\newline
[11] Shorey, A., Cochet, P., Huffman, A., Keech, J., Lueck, M., Pollard, S., and Ruhmer, K., 2014, "Advancements in Fabrication of Glass Interposers," 2014 IEEE 64th Electronic Components and Technology Conference (ECTC), IEEE.
\newline
[12] Attota, R., Dixson, R. G., and Vladár, A. E., 2011, "Through-Focus Scanning Optical Microscopy," Scanning Microscopies 2011: Advanced Microscopy Technologies for Defense, Homeland Security, Forensic, Life, Environmental, and Industrial Sciences, Vol. 8036, SPIE.
\newline
[13] Lu, K., Wang, Z., Chun, H., and Lee, C., 2024, "Wafer Edge Metrology and Inspection Technique Using Curved-Edge Diffractive Fringe Pattern Analysis," Journal of Manufacturing Science and Engineering, 146(7).
\newline
[14] Vartanian, V., Attota, R., Park, H., Orji, G., and Allen, R. A., 2013, "TSV Reveal Height and Dimension Metrology by the TSOM Method," Metrology, Inspection, and Process Control for Microlithography XXVII, Vol. 8681, SPIE.
\newline
[15] Woodham, R. J., 1980, "Photometric Method for Determining Surface Orientation from Multiple Images," Optical Engineering, 19(1), pp. 139-144.
\end{document}